\documentclass[conference]{IEEEtran}

\usepackage{wrapfig}

\usepackage[pdftex]{graphicx}

\usepackage{xcolor}
\ifCLASSOPTIONcompsoc
  \usepackage[caption=false,font=normalsize,labelfont=sf,textfont=sf]{subfig}
\else
  \usepackage[caption=false,font=footnotesize]{subfig}

\ifCLASSINFOpdf

\else

\fi

\usepackage{amsmath}

\begin{document}

\title{Board-level Code-Modulated Embedded Test and Calibration of an X-band Phased-Array Transceiver}

\author{\IEEEauthorblockN{Zhangjie Hong\textsuperscript{1}, Simon Sch{\"o}nherr\textsuperscript{1,2}, Vikas Chauhan\textsuperscript{1}, and Brian Floyd\textsuperscript{1}}

\IEEEauthorblockA{1. North Carolina State University, Raleigh, NC, USA;~~2. Karlsruhe Institute of Technology, Karlsruhe, Germany\\
Email: zhong3@ncsu.edu, sschonh@ncsu.edu, vchauha@ncsu.edu, and bafloyd@ncsu.edu}}

\maketitle

\begin{abstract}
We present methods for built-in test and calibration of phased arrays using code-modulated embedded test (CoMET). Our approach employs Cartesian modulation of test signals within each element using existing phase shifters, combining of these signals into an aggregate code-multiplexed response, downconversion and creation of code-modulated element-to-element ``interference products" using a built-in power detector, demodulation of correlations from the digitized interference response, and extraction of amplitude and phase per element using an equation solver. Rotated-axis methodology is discussed for accurate extraction of phase near the original 0/90/180/270 degree axes. Our techniques are demonstrated at board level for both receive and transmit modes using an eight-element 8-16 GHz phased array constructed using ADAR1000 chips from ADI. At 6 GHz, CoMET-extracted gain and phase are accurate to within 0.2 dB and 3$^{\circ}$ compared to network-analyzer measurements. We then employ CoMET in a calibration loop to determinate optimum control settings at 6 GHz, outside the 8-16 GHz band for which the array was designed. We achieve seven-bit phase resolution with equalized gain. The root-mean squared gain and phase errors are improved from 0.8 dB and 8$^{\circ}$ before calibration to 0.1 dB and 1.7$^{\circ}$ after calibration.
\end{abstract}


\IEEEpeerreviewmaketitle

\begin{figure*}[t]
  \begin{center}
    \includegraphics[width=.9\textwidth]{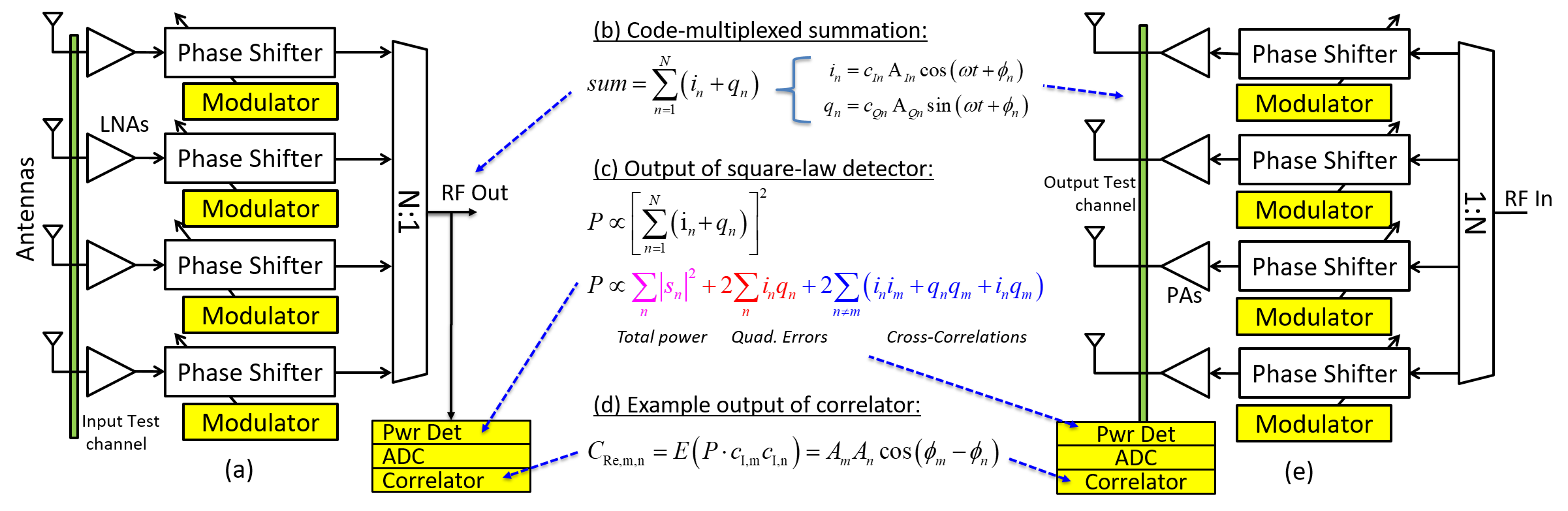}
  \end{center}
  \caption{Block diagrams for (a) receive and (e) transmit arrays, and key equations (b)-(d) for the CoMET methodology.}
  \label{fig:CoMETtheory}
\end{figure*}
\section{Introduction}

Phased arrays typically consist of multiple uniformly-spaced antennas connected to transmit and receive paths which each contain programmable amplitude and phase. Arrays increase antenna gain and  provide the ability to electronically steer and shape a beam. The quality of the array's antenna pattern is directly related to the accuracy of the gain and phase within each element, and this accuracy is compromised by non-idealities such as quadrature error in phase shifters, coupling, temperature and supply gradients, non-orthogonal gain and phase control, and process variations. Furthermore, mismatch in each element's package, interconnect, and antenna can introduce additional errors. Thus, accurate phased-array characterization and calibration techniques with low implementation cost are in high demand.

Built-in self-test (BIST) is a natural solution for this problem and has been used for the characterization and calibration in a variety of radio-frequency (RF) and millimeter-wave (mm-wave) phased arrays \cite{KIM13}-\cite{LIE10}. These phased-array BIST solutions can be classified according to their use of scalar versus vector measurements and their measurement of array elements in a sequential or parallel fashion. 
First,  in-phase (I) and quadrature-phase (Q) downconversion mixers can be used to measure elemental responses in a sequential fashion  \cite{KIM13,KAN16}. The use of IQ mixers provides vector information but requires accurate calibration of the mixer's IQ response before it can be used for BIST. Furthermore, IQ mixers require the use of LO signals which are coherent to the injected test signal. For on-chip BIST, this is straightforward; however, for free-space or package-level test and calibration, this increases complexity.  Second, a ``phasor-sum" approach can be used for BIST, in which two adjacent element's vector signals are combined and fed into a power detector \cite{COH13}--\cite{CHA17}. This technique uses scalar measurements in a sequential fashion. Challenges include the need for multiple calibrated power detectors and an indirect phase measurement which can lead to larger phase error (e.g., 5.9$^\circ$ at Ku band in \cite{CHA17}). Also, this technique is not easy to extend into package and free-space environments.
Finally, code-modulation can be used to test phased arrays, in which orthogonal codes are applied to phase shifters to allow for parallel measurement of the array. In \cite{LIE10}, this is used together with coherent IQ receivers to extract vector information of each element in free space and in parallel. Benefits include a parallel measurement of the array and drawbacks include the need for a coherent, calibrated IQ receiver for BIST.

Code-modulated embedded test (CoMET) is an alternative solution, proposed in \cite{GRE16,GRE18,Comet2019,Comet2021}, which employs a scalar (\textit{i.e.}, power) detector to extract all elemental vectors in a parallel fashion. 
Each element's test signature is modulated using the existing phase shifters and then aggregated into a code-multiplexed response. In contrast to \cite{LIE10}, scalar detection is then used to create a baseband, code-modulated interference response (\textit{i.e.}, interferometry), from which pairwise complex correlations can be extracted using appropriate code-products. These correlations provide information about the full array as well as imperfections introduced during test. Together, these allow extraction of each element's vector response.  

CoMET has multiple advantages. Measurements occur in parallel and in-situ. Test circuit overhead is small, requiring only a means to apply orthogonal modulation, an RF test channel for injecting (or extracting) signals for the antenna ports, and a power detector. Testing overhead is shifted to the digital domain, eliminating the need for large, yield-impacting test circuits, such as an IQ receiver. Finally, the same CoMET infrastructure can be employed for chip, package, and free-space characterization and calibration. Code-modulated interferometry has been used for imaging in \cite{IRIS2016, IRIS2019, IRIS2020, IRIS2021}.

In this paper, we review the CoMET algorithm and discuss a modification we refer to as a ``rotated axis" which improves the phase accuracy near the original 0/90/180/270 Cartesian axes. We also show how CoMET can be used for board-level self-test of an eight-element array, employing two 8--16 GHz packaged phased arrays from Analog Devices.  Finally, we present a calibration method for the array in which CoMET is used within a closed loop to equalize gain response across the array and achieve an optimized seven-bit phase response in hardware.

\section{CoMET Methodology}
\subsection{Review of Original Approach}
We first review the CoMET algorithm presented in \cite{GRE18}.  Our approach is illustrated in Fig. \ref{fig:CoMETtheory} along with key equations. 
\begin{enumerate}
\item A test signal is injected into the array. For transmit (TX) arrays, this injection uses the normal signal path. For chip- and board-level receiver (RX) test, this injection requires a built-in test channel which weakly couples a signal into each element. For free-space RX test, this injection can be over-the-air. 
\item Cartesian modulation is applied to each phase shifter to uniquely encode both the I and Q response in each element. Two degrees of freedom are asserted to allow extraction of the vector in Cartesian form. This encoding results in the phase being modulated across four quadrants; hence, four-quadrant information is obtained within a single CoMET test frame.  
\item All elemental signals are combined (Fig. \ref{fig:CoMETtheory}(b)) to form a code-multiplexed test response. For an RX array, this aggregation uses the normal signal path, whereas for a TX array, this aggregation network would have the same topology as the RX array's injection network.  
\item The code-multiplexed summation is squared using a scalar (power) detector to generate a low-frequency interference response (Fig. \ref{fig:CoMETtheory}(c)). This interferometric technique produces a signal which contains all possible correlations (II, QQ, and IQ) between elemental signals. For an N-element array, the number of correlations is $C_2^{2N}={2N(2N-1)}/{2}$.  Each correlation is modulated according to a code product; hence, the original codes used within step two are chosen to have orthogonal code products (OCP). 
\item Each correlation is demodulated by correlating the squared summation with the OCP of interest (Fig. \ref{fig:CoMETtheory}(d)). These include intra-element IQ correlations which provide information on quadrature error within each phase shifter and inter-element IQ correlations which provide information on phase offset introduced within the test-signal injection network.
\item All correlations are used within an equation solver from which we can extract 2N individual amplitudes (I and Q) and 2N individual phase responses (I and Q, incorporating phase offset) for all elements. 
\end{enumerate}

In \cite{GRE18}, CoMET is validated using a 60 GHz four-element receive array. BIST measurement for a four-bit phase response show CoMET functions properly and extracts gain and phase with 1 dB and 4$^{\circ}$ accuracy with respect to a vector network analyzer (VNA).  These errors, though, are larger then desired for a higher resolution phase shifter measurement. Here, we discuss a rotated axis methodology to address this issue. Additionally, assumptions were made in \cite{GRE18} about the injected phase offset between elements being below 10$^{\circ}$, allowing us to neglect it for II and QQ correlations. Here, we employ CoMET on a board-level system which has larger injected phase offset. As a result, our equation solver is updated to fully incorporate this phase offset into all correlations.

\begin{figure}
 \centering
	\subfloat[]{\includegraphics[width=0.18\textwidth]{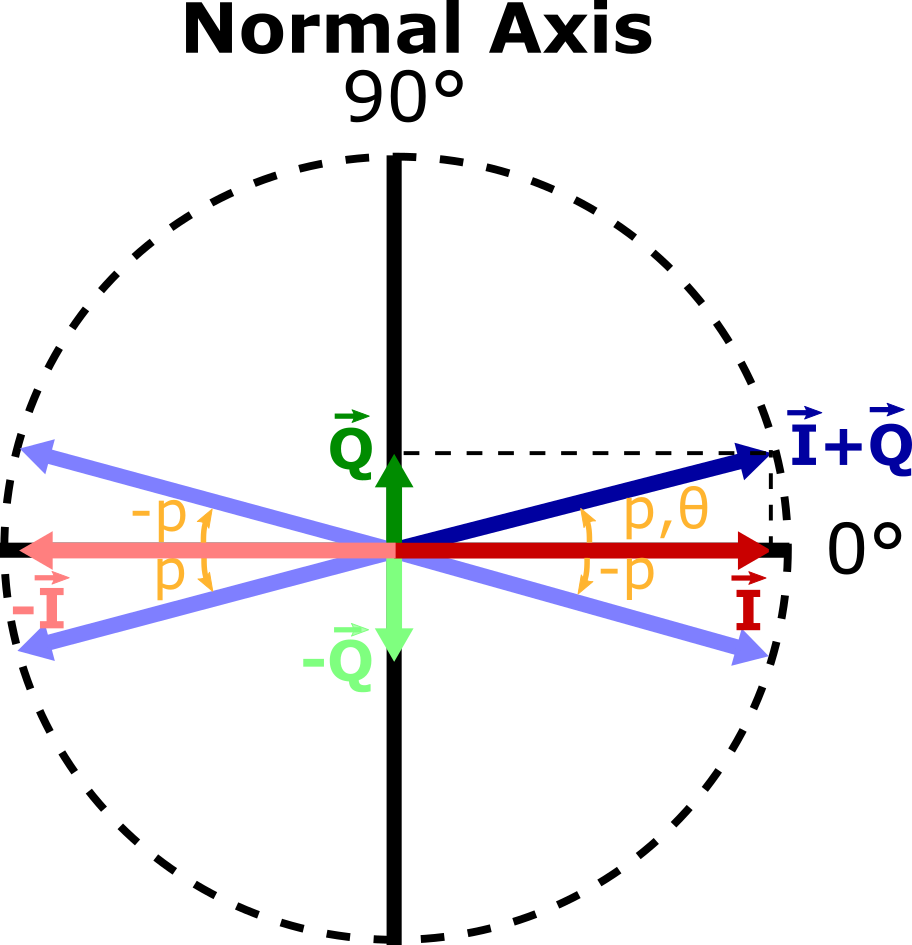}
		\label{fig:ModNorm}}
	\subfloat[]{\includegraphics[width=0.22\textwidth]{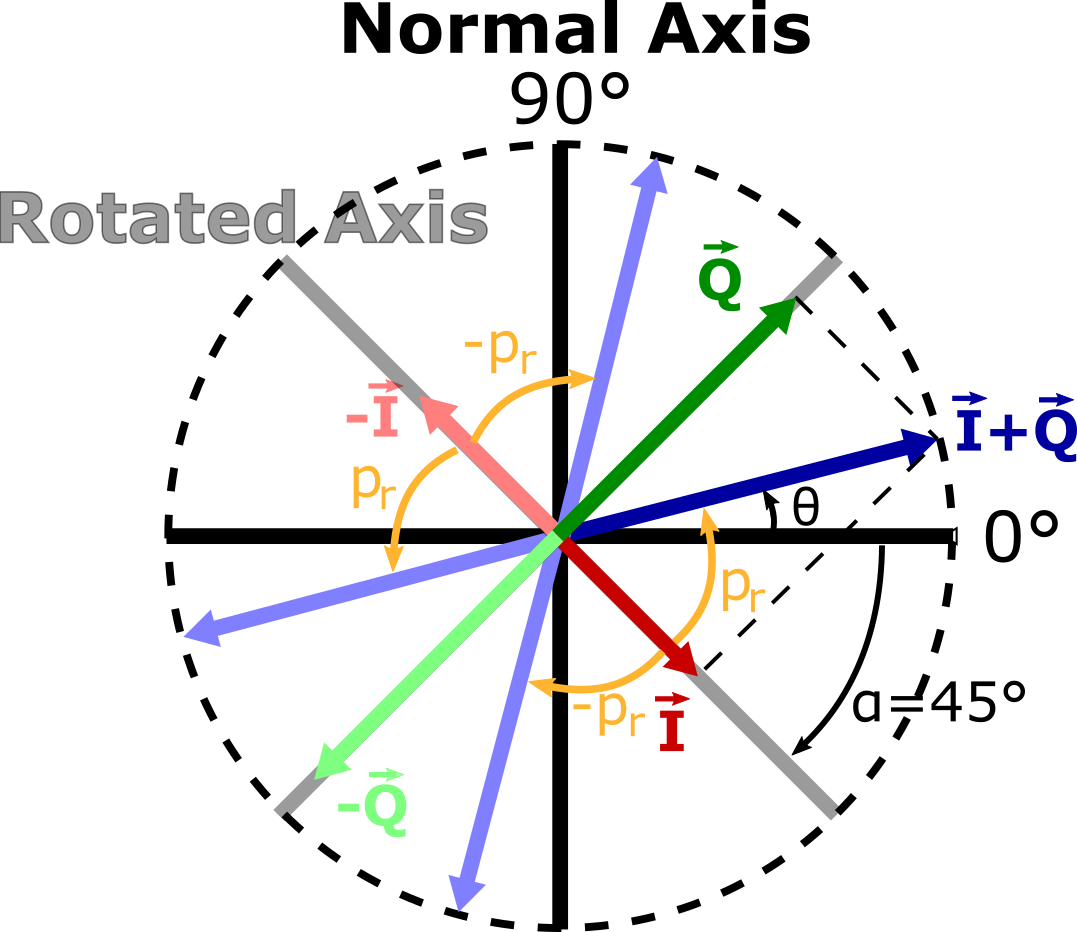}
		\label{fig:ModRot}}
	\caption{The vector-representation of modulated signals in (a) normal axis, and (b) rotated axis.}
	\label{fig:Modulation}
\end{figure}
\begin{figure}
  \begin{center}
    \includegraphics[width=0.3\textwidth]{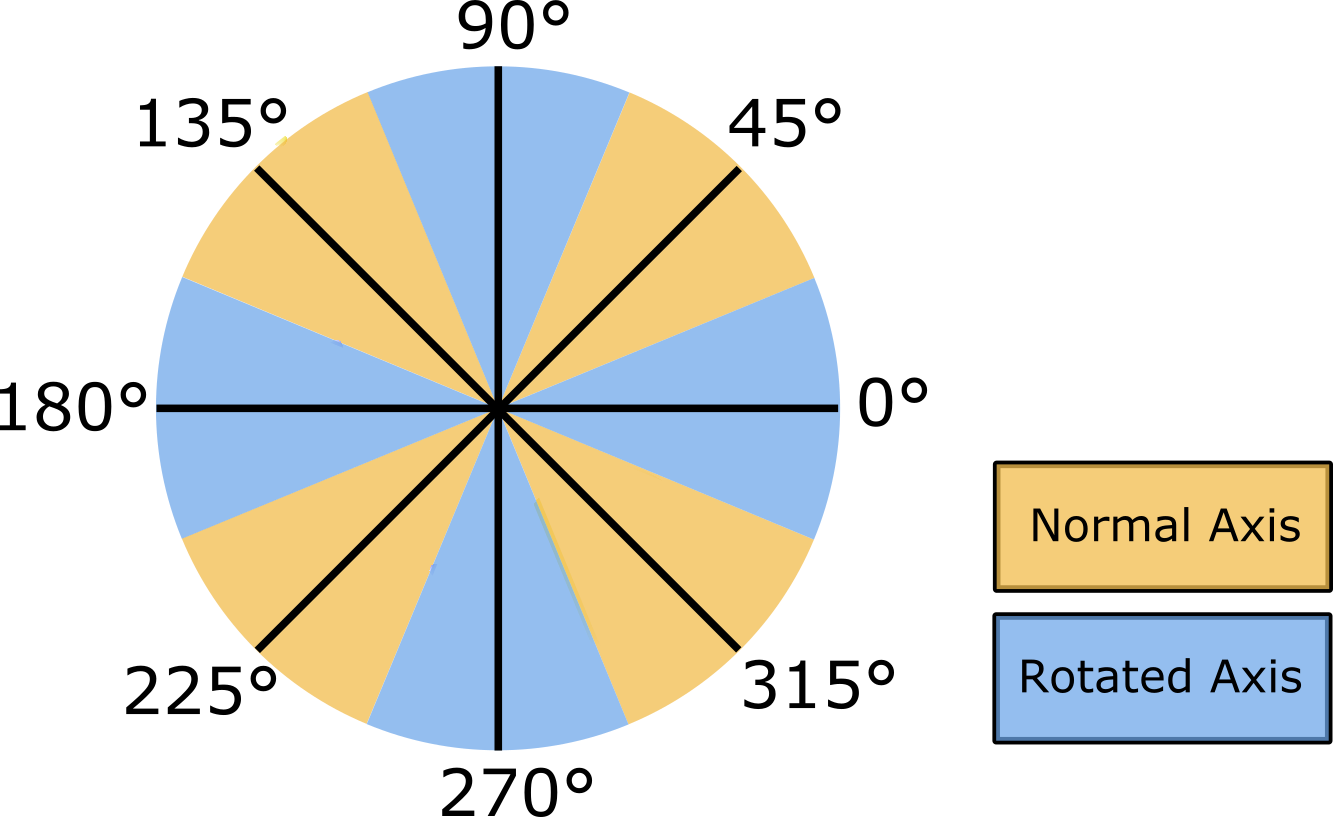}
  \end{center}
  \caption{The usage division of normal axis and rotated axis}
  \label{fig:RotatedAxes}
\end{figure}
\subsection{Rotated axis}
As stated above, the amplitude and phase of each path is obtained by demodulating all cross-correlations from the interference response and then solving the system of cross-correlation equations. Noise is present within each demodulated correlation due to the other orthogonal code-modulated terms as well as circuit noise. When the phase is close to the original Cartesian axes, the amplitude of either I or Q vector is close to zero; thus, the relevant cross-correlations will have a low signal-to-noise ratio.  As a result, amplitude and phase error increases as the phase approaches the Cartesian axes.

To address this problem, the modulation axes can be ``rotated" by $\alpha=45^{\circ}$ when the phase is close to the original quadrant boundaries. This corresponds to modulating a rotated set of basis functions for the phase shifter. For a given phase shift, equivalent expressions within either axis system are
\begin{equation}
\begin{split}
    E_n\left ( t \right )&=A_{I,n} cos\left ( \omega t  \right)-A_{Q,n} sin\left ( \omega t  \right )\\
    &=A_{Ir,n}cos\left ( \omega t +\alpha \right )- A_{Qr,n}sin\left ( \omega t +\alpha \right )
\end{split}
\end{equation}
 The first line corresponds to the original IQ axes, where $A_{I,n}=A_n cos\left (\theta  \right )$, $A_{Q,n}=A_n sin\left (\theta  \right )$. The second line corresponds to the rotated IQ axes, where $A_{Ir,n}=A_n cos\left (\theta -\alpha \right )$, $A_{Qr,n}=A_n sin\left (\theta -\alpha \right )$. 
 
In the original axis system, amplitudes $A_{I,n}$ and $A_{Q,n}$ are modulated by the in-phase and quadrature-phase codes resulting in the four following ideal phase states: $\theta=\pm p$, $180\pm p$, where $p=tan^{-1}|A_{Q,n}/A_{I,n}|$. In the rotated axis system, amplitudes $A_{Ir,n}$ and $A_{Qr,n}$ are  modulated, resulting in the four following ideal phase states: $\theta=\alpha\pm p_r$, $180+\alpha\pm p_r$, where $p_r=tan^{-1}|A_{Qr,n}/A_{Ir,n}|$, as shown in Fig. \ref{fig:Modulation}.

In this work, we propose the overall unit circle to be partitioned as shown in Fig. \ref{fig:RotatedAxes}. The original axes are used to measure phase states close to the rotated axes (45$^{\circ}$/135$^{\circ}$/225$^{\circ}$/315$^{\circ}$) and the rotated axes are used to measure states close to the original axes (0$^{\circ}$/90$^{\circ}$/180$^{\circ}$/270$^{\circ}$).
In each portion, the measured phase is at least 22.5$^{\circ}$ away from the applied axis, producing sizeable Cartesian vectors across phase settings and improving measurement accuracy.

For both original and rotated axes, the original CoMET algorithm can be used to extract either $p$ or $p_r$ (while also including the impact of quadrature offset and injection offset as in \cite{GRE18}).  What is needed is an accurate estimate for the axis rotation, $\alpha$.  Ideally, this should be 45$^{\circ}$; however, circuit non-idealities, namely quadrature error, can result in an offset. We therefore set the phase shifter to a 0$^{\circ}$ setting and extract the phase in the rotated axis. This equals $\alpha$.

\begin{figure}
	\centering
	\subfloat[]{\includegraphics[width=0.35\textwidth]{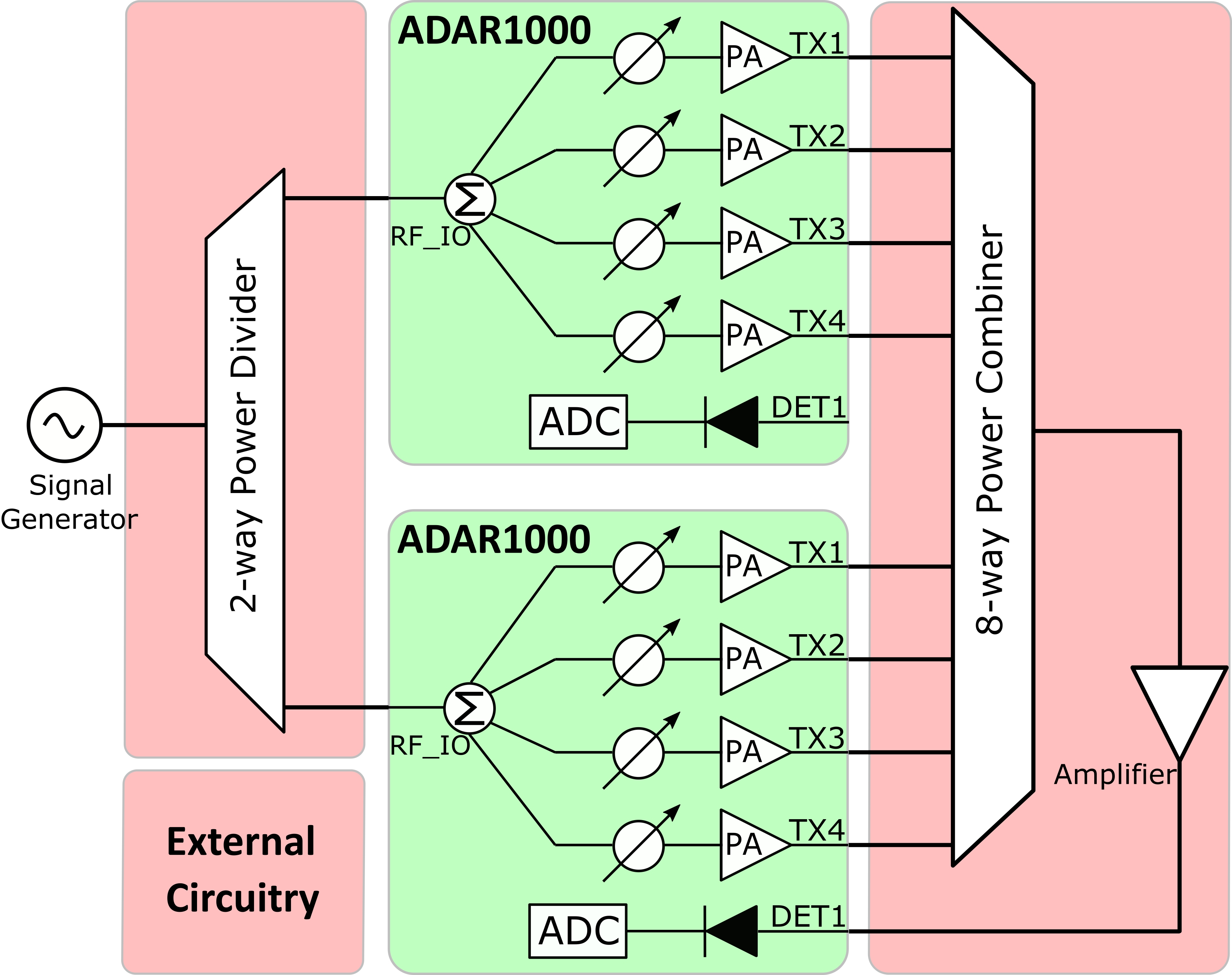}
		\label{fig:Block_Diagram_RX}} \\
	\subfloat[]{\includegraphics[width=0.35\textwidth]{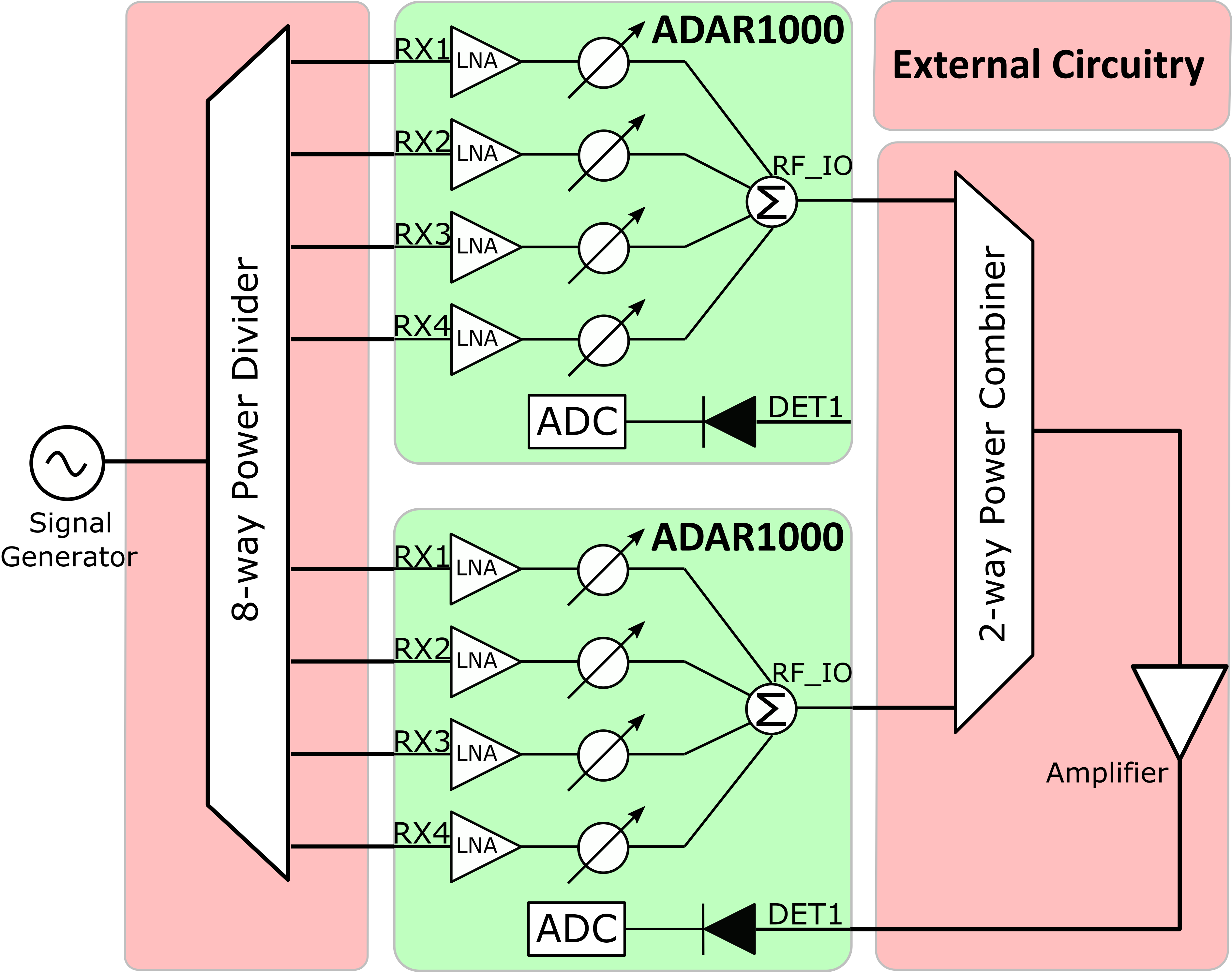}
		\label{fig:Block_Diagram_TX}}
	\caption{Block diagrams of the (a) eight-element TX array test setup, and (b) eight-element RX array test setup.
	}
	\label{fig:interferometer}
\end{figure}

\section{Board-Level CoMET Experiments}
\subsection{Hardware Description}
To evaluate CoMET for our new algorithm in both receive and transmit modes, we use the Analog Devices ADAR1000 8--16GHz phased-array \cite{ANA181}. 
Each ADAR1000 chip contains four identical TX and RX channels for time division duplex operation. An on-chip four-way power combiner/splitter generates a sub-array input/output (RF\textunderscore IO).  
In each TX/RX channel, there is a power amplifier (PA)/low noise amplifier (LNA), a variable-gain amplifier (VGA), a switchable attenuator and a phase shifter. The VGA and the switchable attenuator together provide a 31 dB tuning range of channel gain.  
The phase shifter is implemented with a vector interpolator topology, using I- and Q- VGAs. 
--Six bits control each VGA- five bits for amplitude control and one bit for polarity control- for a total of 12 bits per phase shifter. The polarity control bit is used to apply the CoMET modulation for the I and Q vectors. 
Finally, four on-chip power detectors (PD) are included along with an analog-to-digital converter (ADC), which digitizes the PD output.

\begin{figure}
 \centering
	\includegraphics[width=0.4\textwidth]{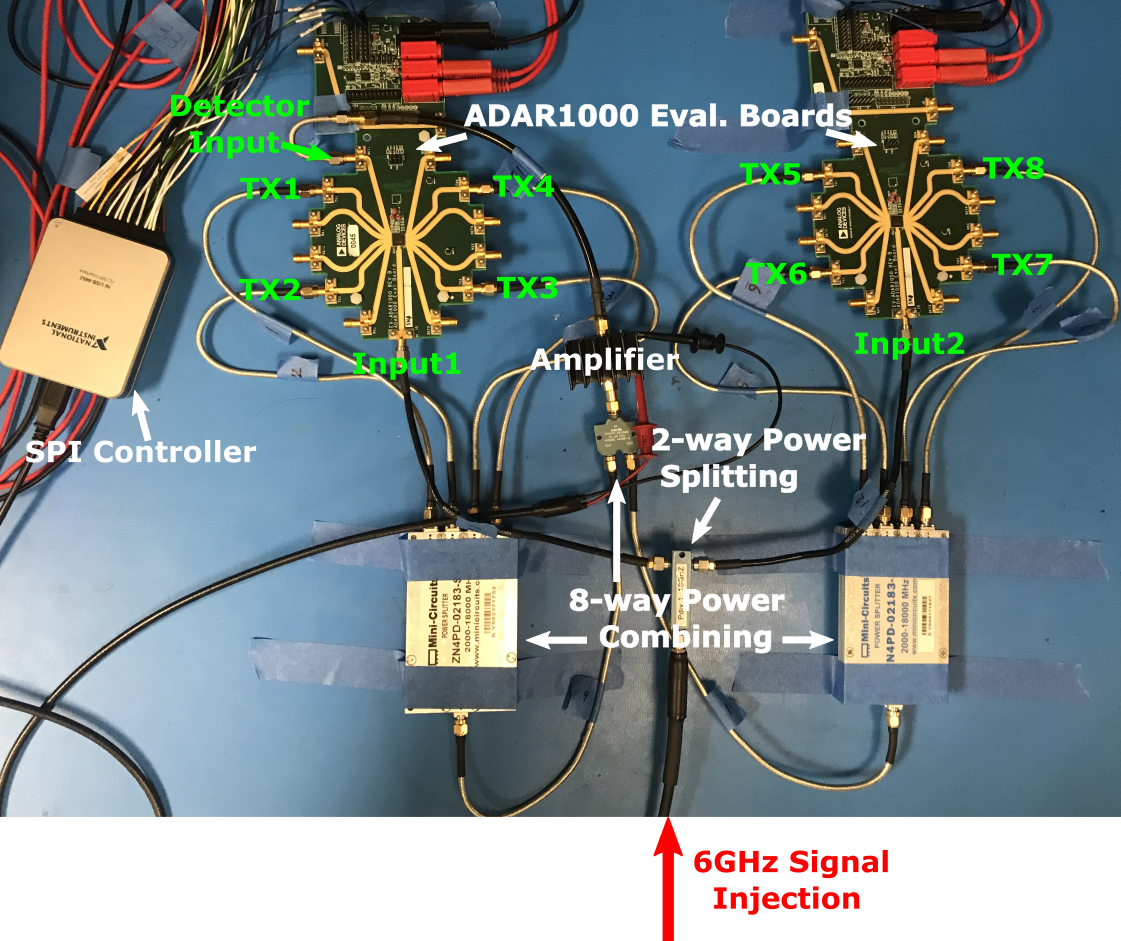} \label{fig:Photo_TX}
	\caption{Photograph of the eight-element TX array test setup. }
	\label{fig:photo}
\end{figure}

In our experiments, an eight-element array is created using two separate ADAR1000 evaluation boards \cite{ANA182}. Block diagrams of the TX and RX test setup are shown in Fig. \ref{fig:interferometer} and a photograph of the TX setup in Fig. \ref{fig:photo}. An RF input test signal is generated using a signal generator. The injection networks for CoMET are created using a coaxial two-way splitter for the TX and a coaxial eight-way splitter for the RX. Conversely, the CoMET extraction network is formed using an eight-way combiner for the TX and a two-way combiner for the RX. The combined signal is then fed into an external amplifier to increase power level and then back into one power detector located on an ADAR1000. 
All modulations are applied using the serial interface and all scalar power readings are obtained through the on-chip ADC, whose values are read out through the serial interface. 
Sixteen 256-bit long Walsh-Hadamard codes are allocated to the eight I vectors and eight Q vector. Codes are selected to have orthogonal code products (OCP).

\begin{figure*}
 \centering
	\subfloat[]{\includegraphics[width=0.33\textwidth]{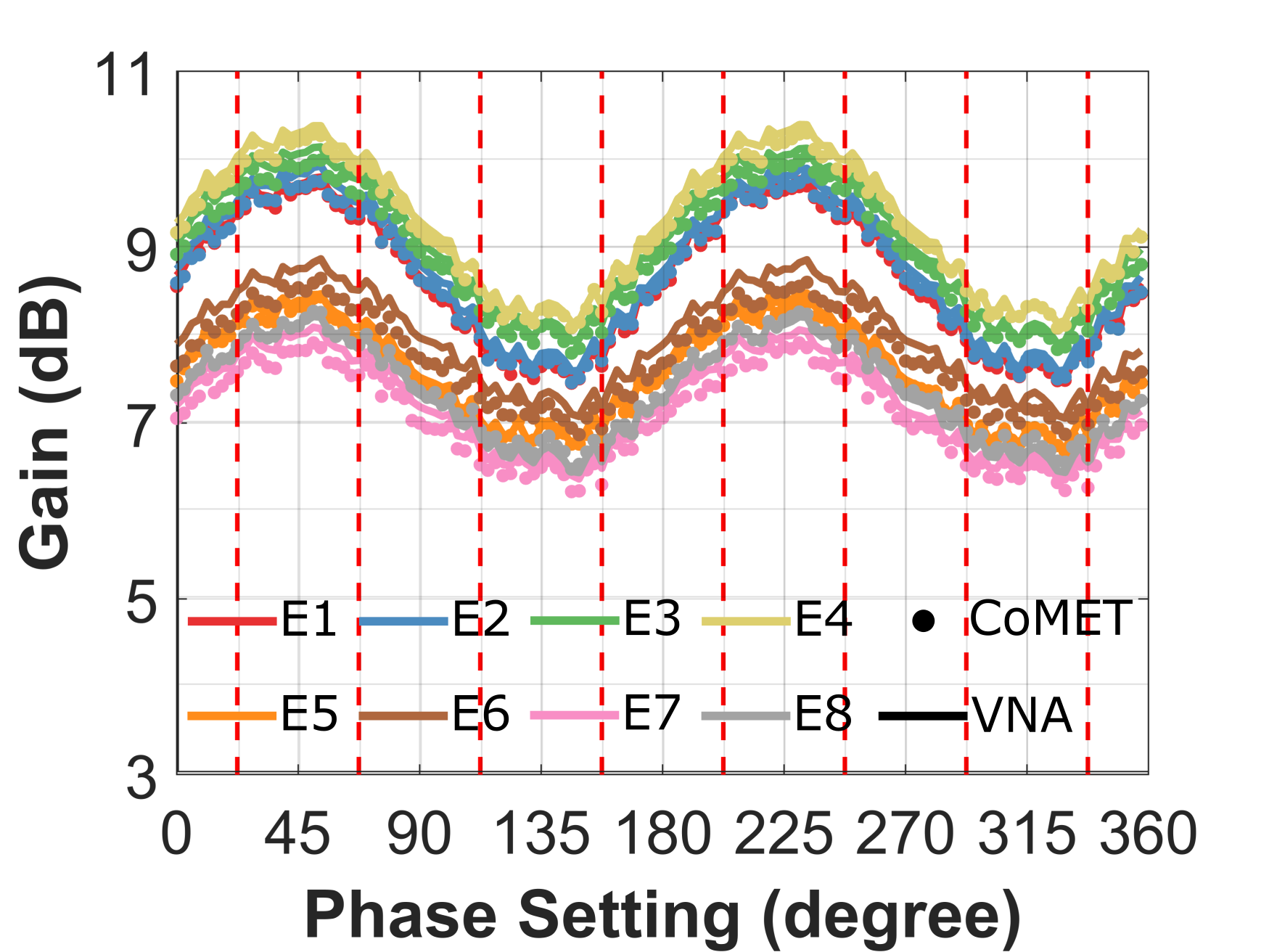}
		\label{fig:EXGain18_TX}}
	\subfloat[]{\includegraphics[width=0.33\textwidth]{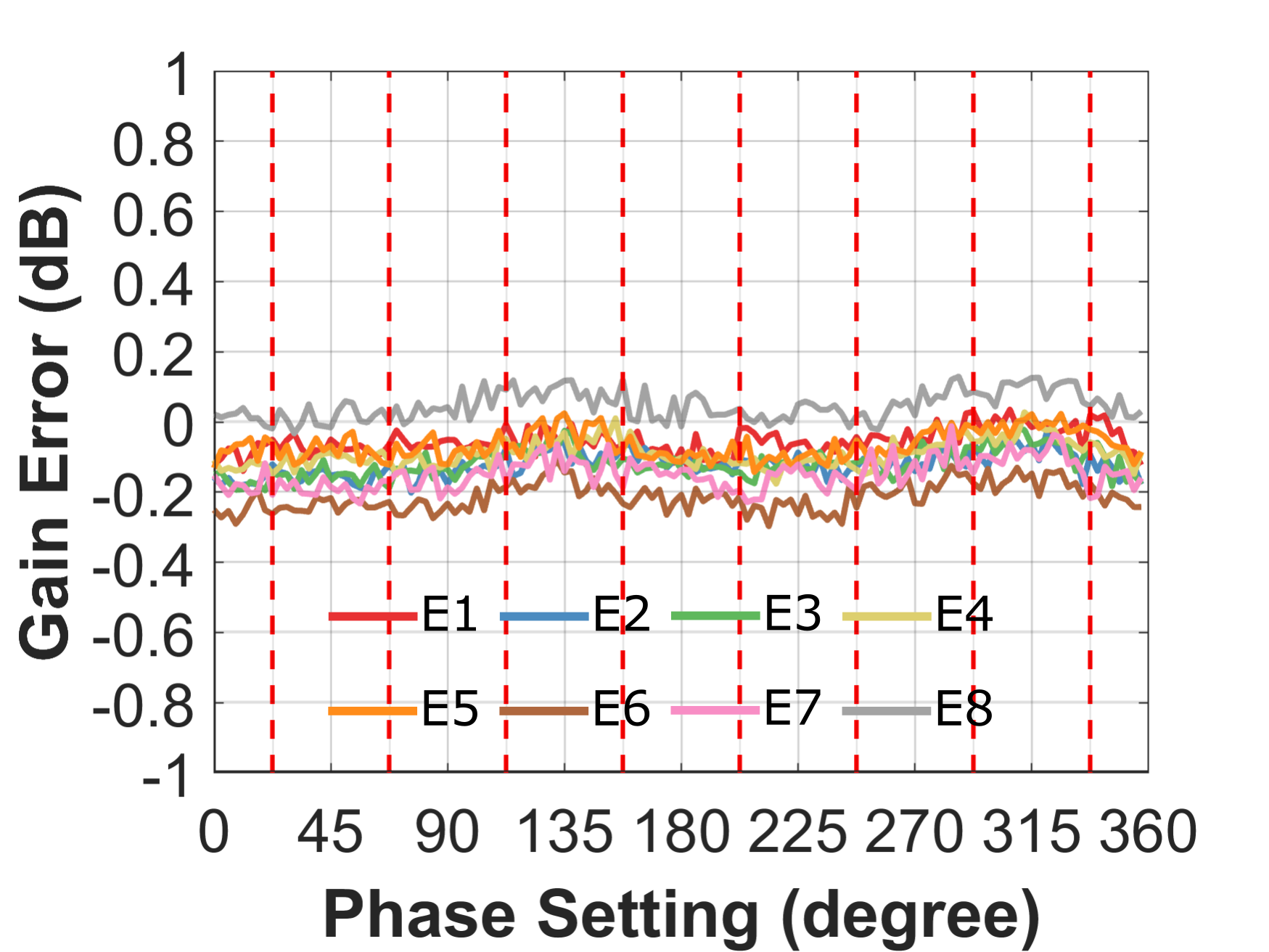}
		\label{fig:EXGainError_TX}}
	\subfloat[]{\includegraphics[width=0.33\textwidth]{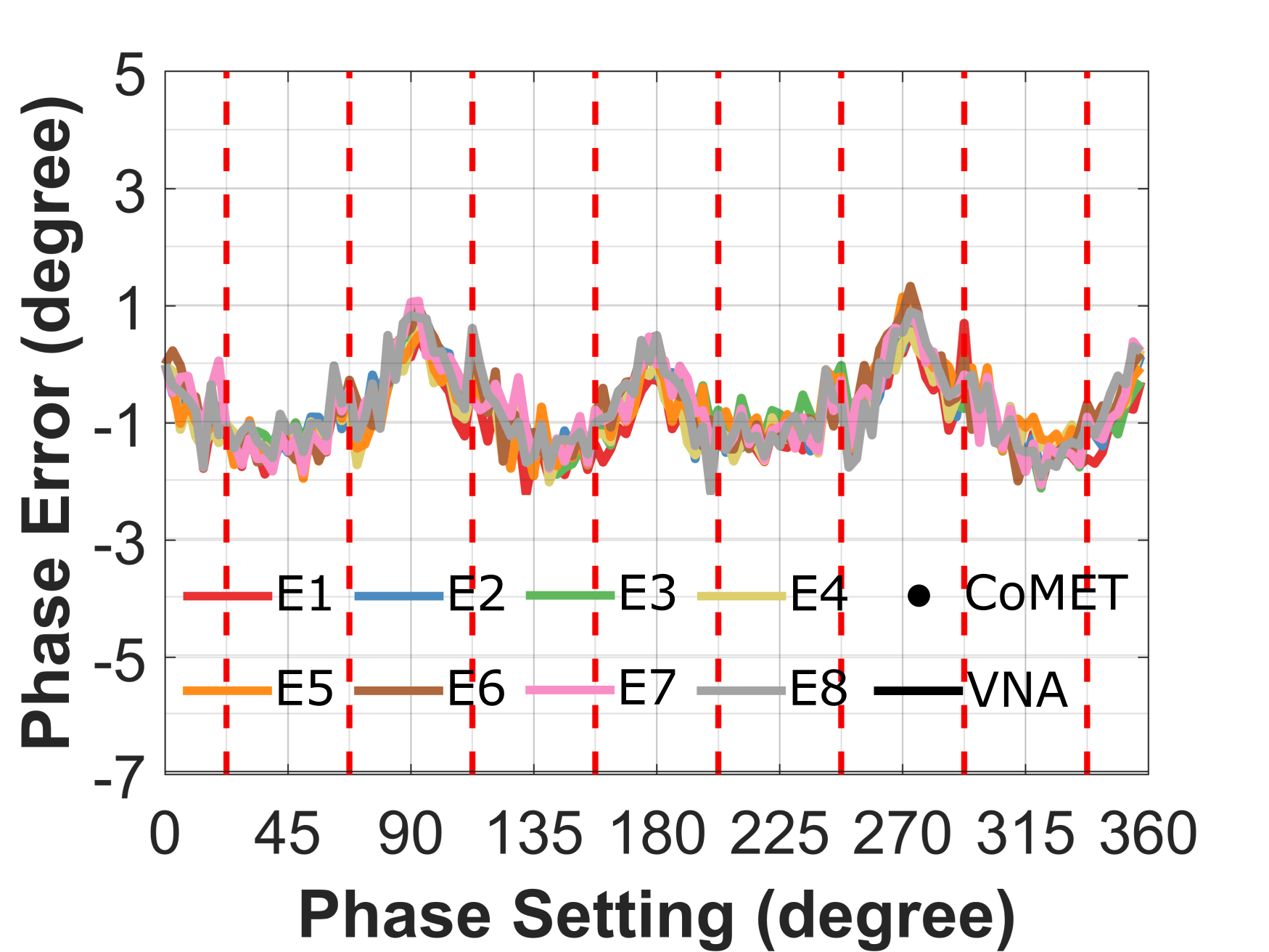}
		\label{fig:EXPhase18_TX}}
	\caption{Comparison between VNA measurements and CoMET extractions in TX mode: (a) gain response, and (b) gain error; (c) phase error. 
	}
	\label{fig:EXGainTX}
\end{figure*}

\begin{figure*}
 \centering
	\subfloat[]{\includegraphics[width=0.33\textwidth]{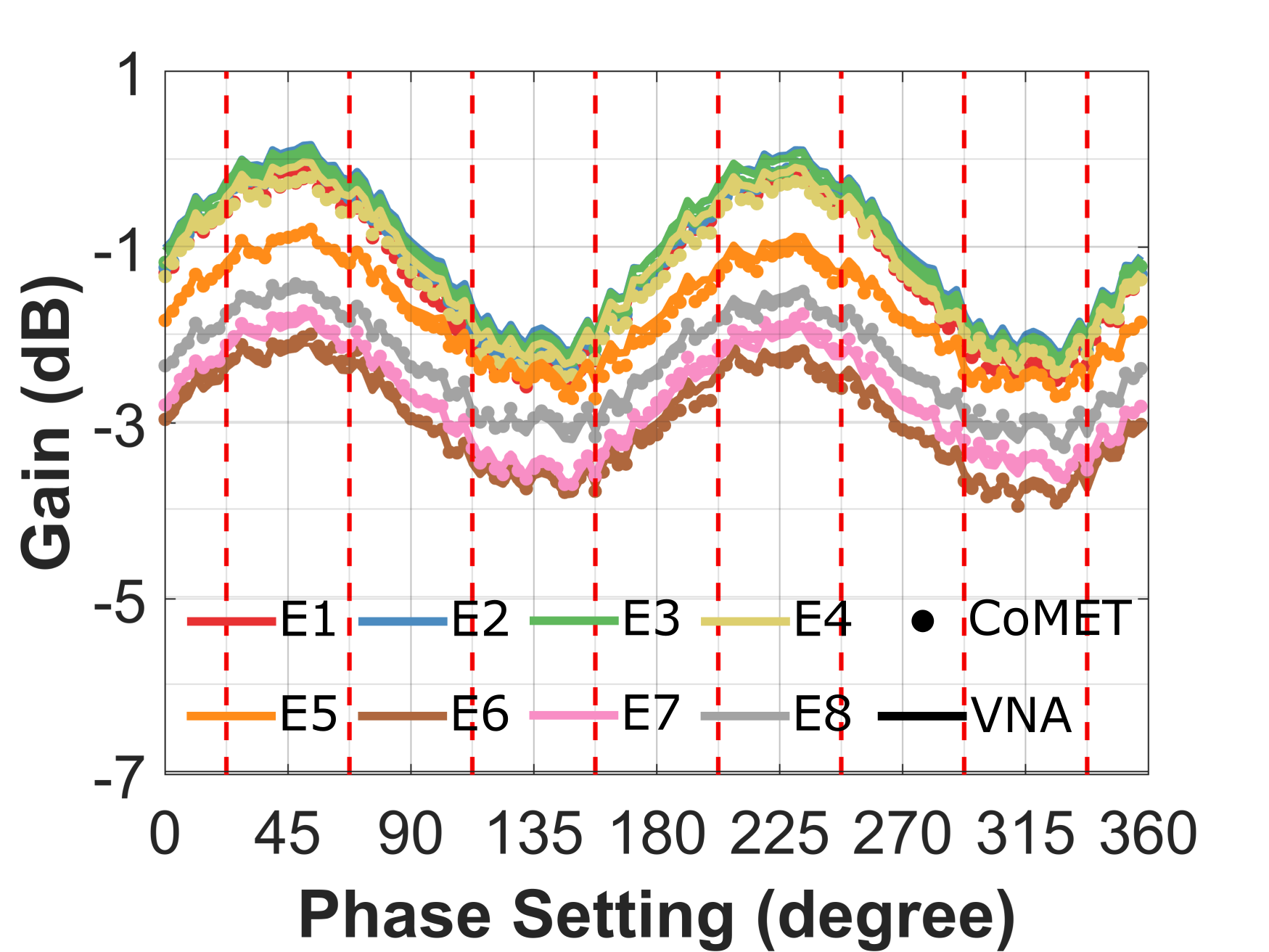}
		\label{fig:EXGain18_RX}}
	\subfloat[]{\includegraphics[width=0.33\textwidth]{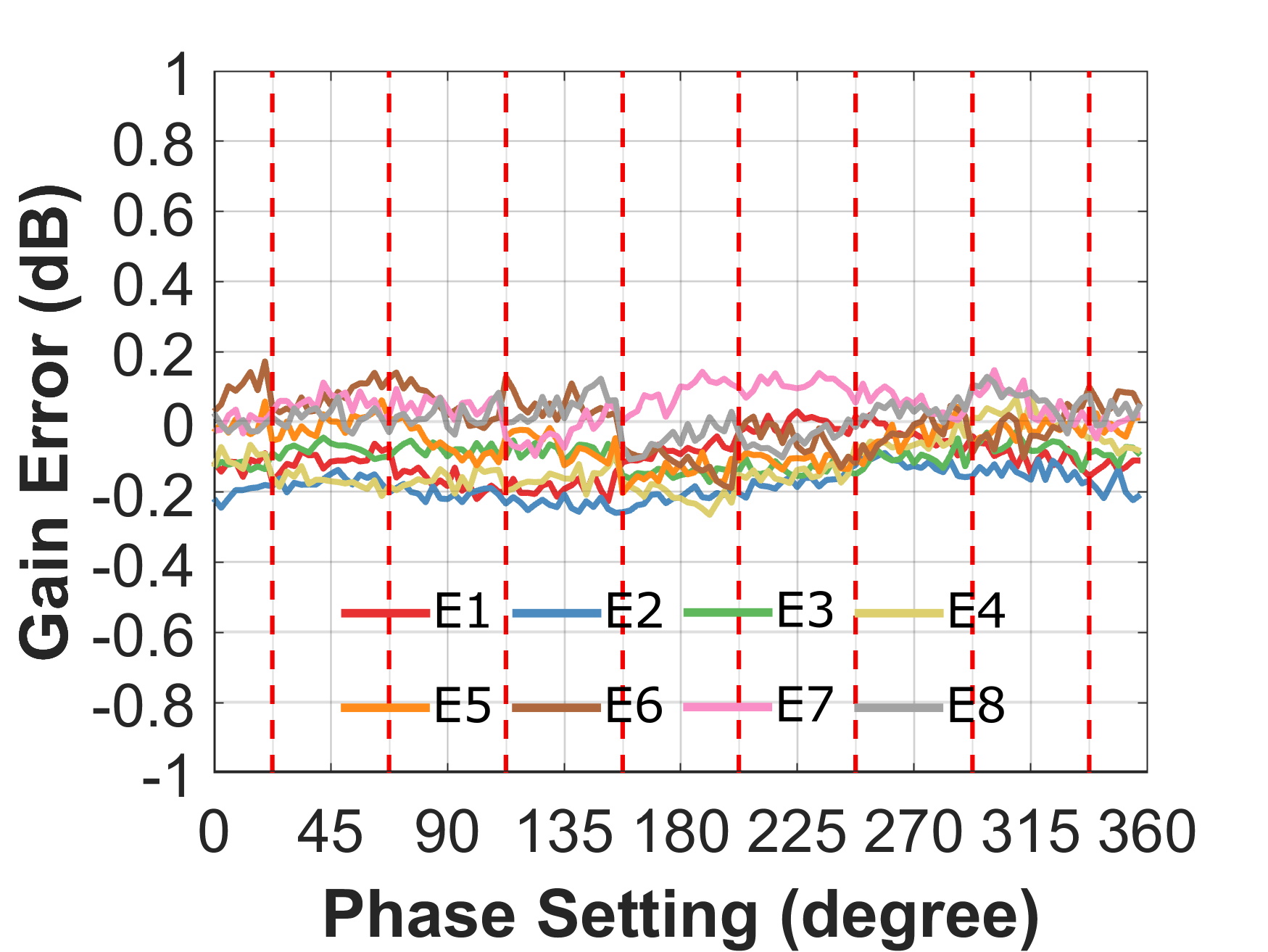}
		\label{fig:EXGainError_RX}}
	\subfloat[]{\includegraphics[width=0.33\textwidth]{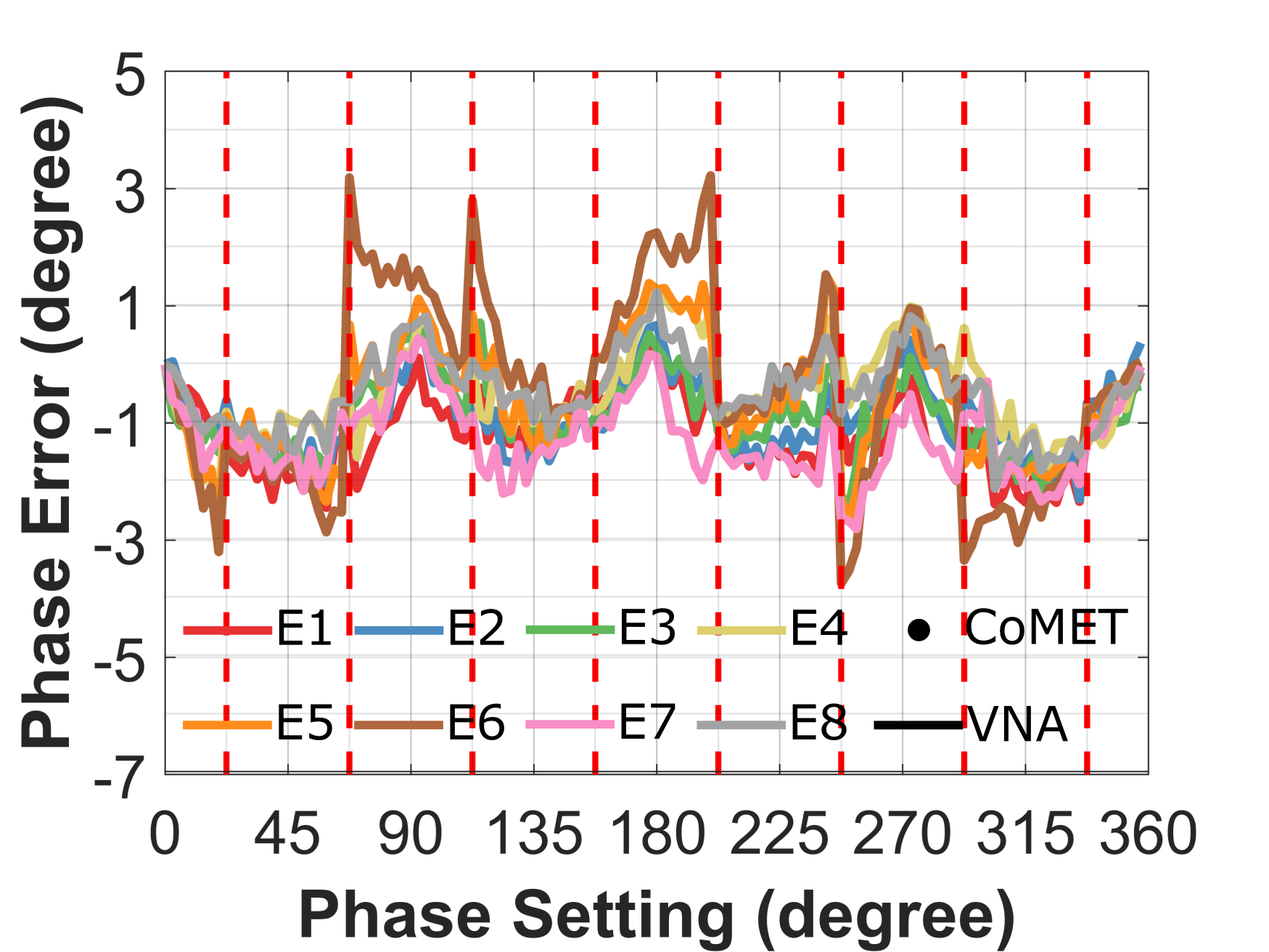}
		\label{fig:EXPhase18_RX}}
	\caption{Comparison between VNA measurements and CoMET extractions in RX mode: (a) gain response, and (b) gain error; (c) phase error.  
	}
	\label{fig:EXGainRX}
\end{figure*}

\subsection{Transmitter Results}

First, a VNA is used to measure all eight elements of the array, with reference planes located at the input of the injection network and the output of the external amplifier. The s-parameters are taken across all phase settings, with values selected according to the manufacturer look-up table (LUT). This LUT had been optimized for 8 GHz. Also, the s-parameters are taken with all eight elements active, leading to the array having the same temperature and power supply drop which is present when the full phased array is active. This s-parameter data serves as our reference data.

Second, CoMET is run for the array using the methods described previously. Note that CoMET uses the same reference plane as the VNA. For each phase shift, a 256-bit modulation is applied and for each modulation state, the digitized power detector (PD) reading is read through the serial interface. The phase is then swept across a single quadrant, where CoMET measures four quadrant information within this sweep. For seven-bit phase resolution, we have 32 states per quadrant; thus, the total number of modulation states and PD readings for full seven-bit CoMET phase extraction is 8192.

The gain and phase response across phase settings from CoMET extraction are compared to the results from VNA measurement in Fig. \ref{fig:EXGainTX}. For our demonstration, we  measure at 6 GHz which is deliberately chosen below the intended 8--16 GHz frequency range. This is done to demonstrate CoMET calibration capabilities in the next section. The result shows an excellent agreement between CoMET extraction and VNA measurement, where the gain and phase match within 0.2 dB and 2$^\circ$ across phase settings. This is considerably better than our prior work in \cite{GRE18}.

\subsection{Receiver Results}
In RX mode, a similar experiment setup to that of the TX is used, except the injection and extraction networks are interchanged. Once again, the RX array is first measured using a VNA and then measured using CoMET at 6 GHz. Fig. \ref{fig:EXGainRX} compares the gain and phase response across phase settings from CoMET extraction and the VNA measurement. Once again, the CoMET extraction agrees well with the VNA measurement, where the gain and phase match within 0.2 dB and 3$^\circ$ across phase settings.

\section{Closed-Loop CoMET Calibration}
\subsection{Algorithm}
Given CoMET's capabilities, it is possible to embed it within a loop to calibrate the phased array. 
Our calibration goal is to equalize the gain across all eight elements and select optimum phase settings for seven-bit phase resolution (128 states). 
Using CoMET, we can extract the vector response for each element in the array in parallel, and then for each element, calculate an error vector magnitude (EVM). This is the magnitude of the error vector between the extracted vector ($A_{extr}$, $\theta_{extr}$) and the desired vector($A_{des}$, $\theta_{des}$), defined as follows for element $n$.
\begin{equation}
EVM_n =  \sqrt{(A_{I,extr,n}-A_{I,des,n})^2+(A_{Q,extr,n}-A_{Q,des,n})^2}
\end{equation}
Here, the I and Q amplitudes are defined as before, with $A_{I}=Acos(\theta)$ and $A_{Q}=Asin(\theta)$. 

In our work, we aim to minimize EVM for every phase state for all array elements to create an 8x128 look-up table for the array (for a given gain setting). 
The closed-loop phased array calibration follows the steps below:
\begin{enumerate}
\item Start from initial I and Q settings which can be obtained from an estimated LUT or a previously calibrated LUT.
\item Run CoMET to extract gain and phase and calculate EVM for all elements.

\item Evaluate EVM using CoMET for eight adjacent states (\textit{i.e.}, increment/decrement VGAs in the IQ phase shifter). 
\item Repeat steps 2 and 3 above until the EVM of the current state is less than that of any of the eight adjacent states for all eight elements. 

\item Repeat steps 1-4 for all phase settings.
\end{enumerate}

Total operation time is a significant specification for a calibration method,
\begin{equation}
t_{total} = t_{frame} \cdot n_{state} \cdot n_{iteration} \cdot n_{phase}  
\end{equation}
where $t_{frame}$ is the running time of a single CoMET frame, $n_{state}$, $n_{iteration}$, $n_{phase}$, are the number of states per iteration, the average number of iterations per phase setting and the number of phase settings, respectively. CoMET calibration can be accelerated by reducing these terms. For example, we can reduce the number of iterations per phase setting by changing step size based on gradient descent. Reducing overall the running time will be the subject of future work.

\subsection{Calibrated Hardware Results}

\begin{figure}
 \centering
	\subfloat[]{\includegraphics[width=0.35\textwidth]{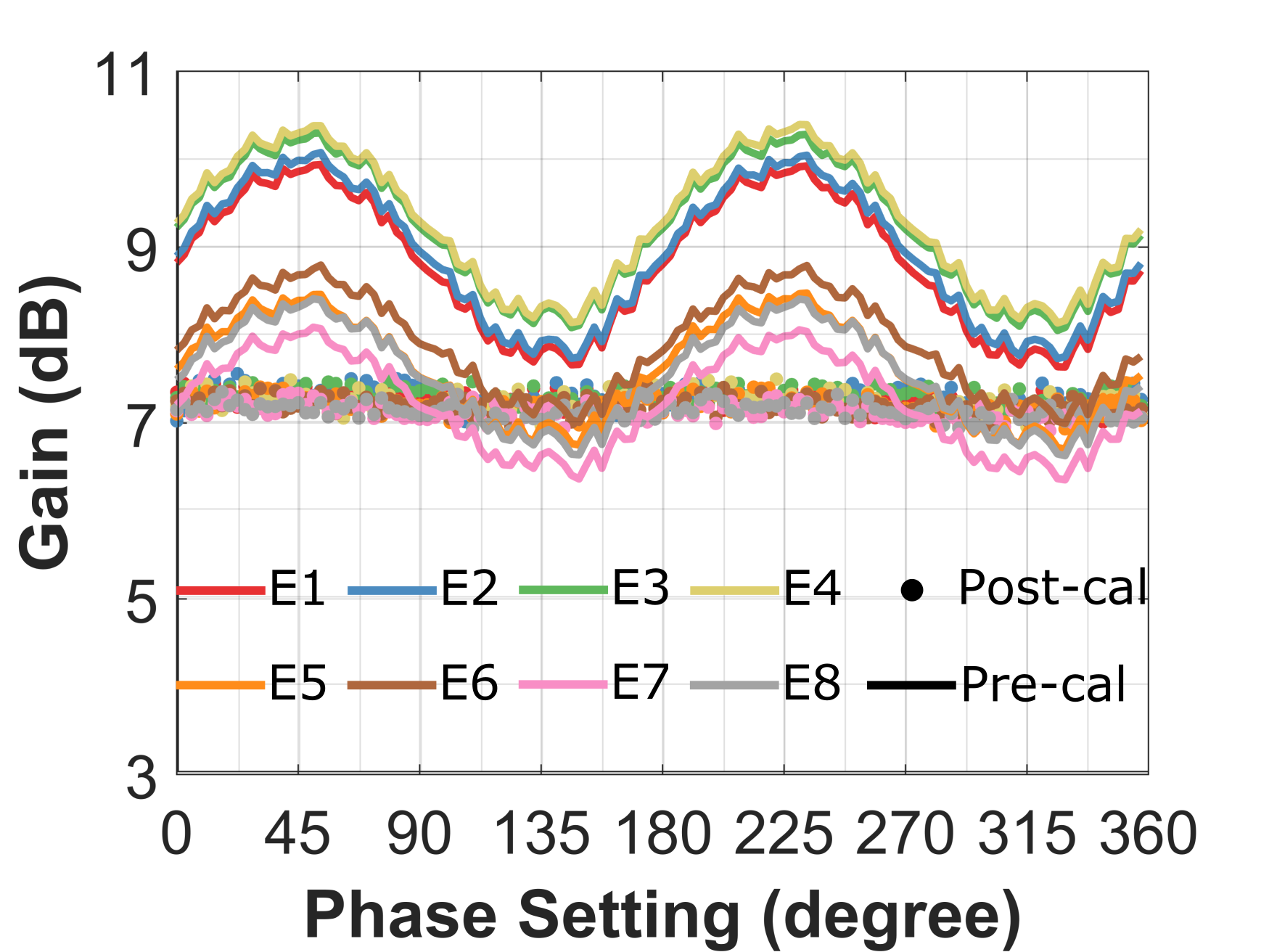}
		\label{fig:POSTGain18_TX}}\\
	\subfloat[]{\includegraphics[width=0.35\textwidth]{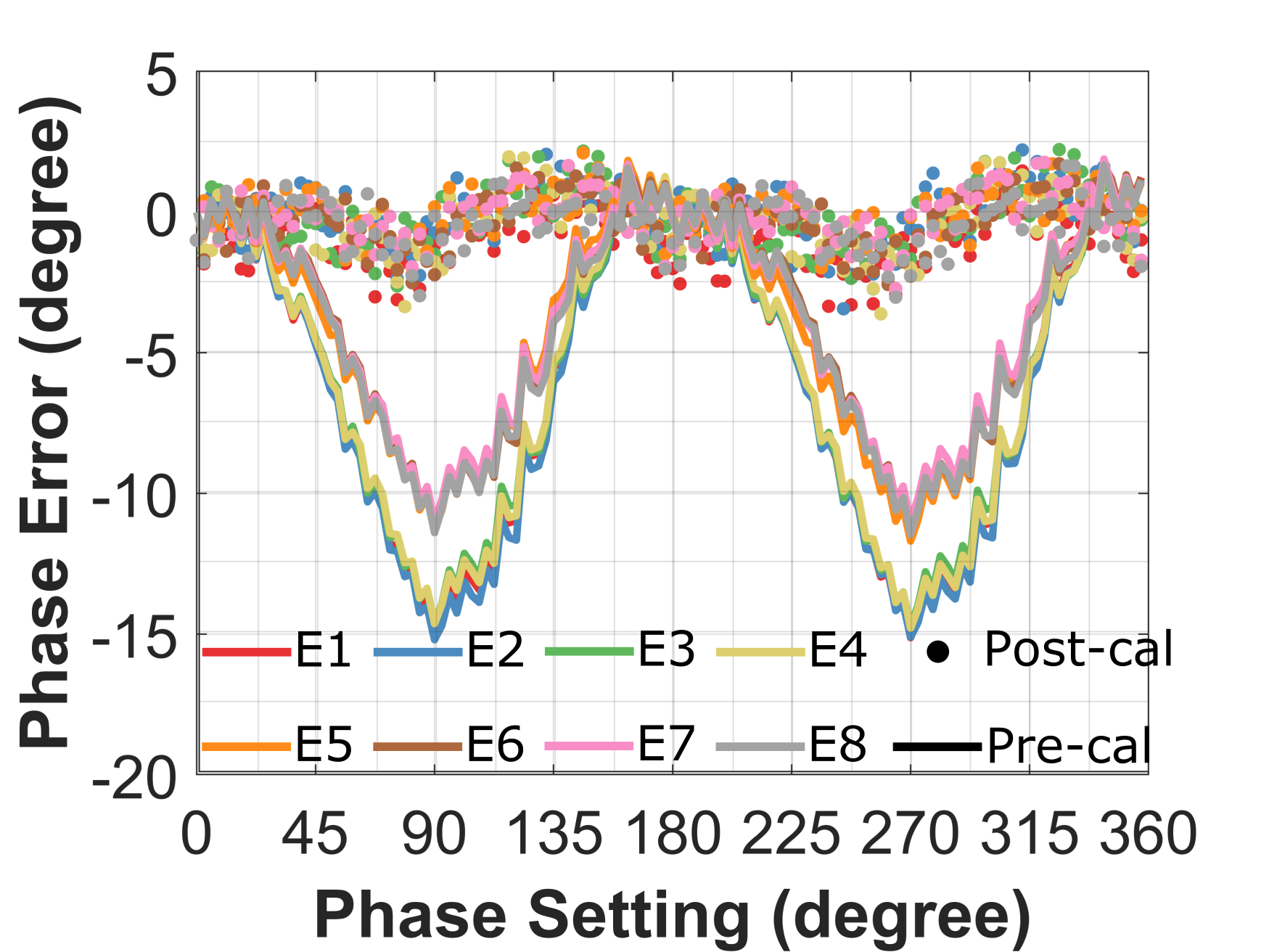}
		\label{fig:POSTPhase18_TX}}

	\caption{VNA measured (a) gain and (b) phase response across phase settings before and after calibration in TX mode.}
	\label{fig:POSTGainTX}
\end{figure}

\begin{figure}
 \centering
	\subfloat[]{\includegraphics[width=0.35\textwidth]{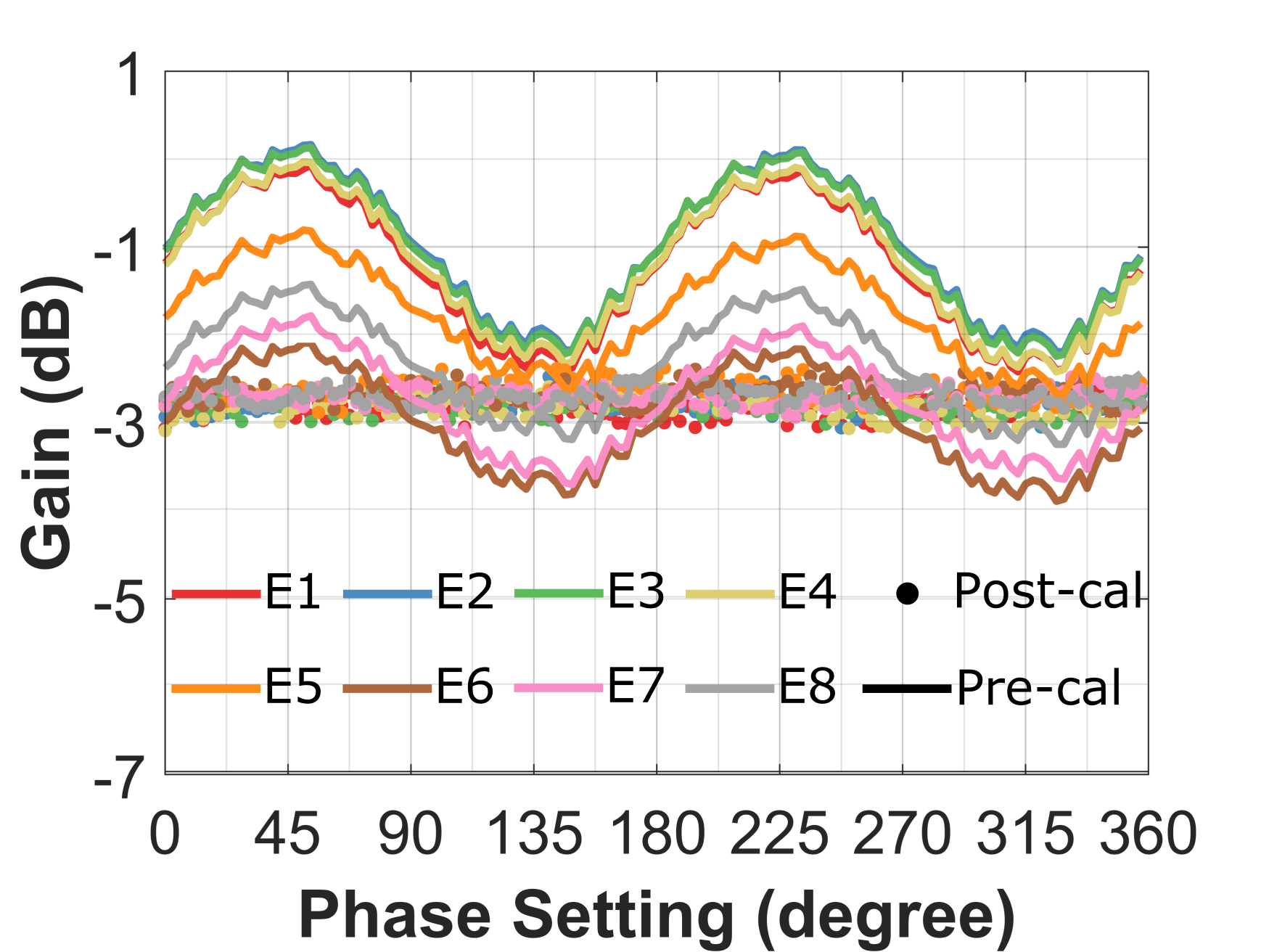}
		\label{fig:POSTGain18_RX}}\\
	\subfloat[]{\includegraphics[width=0.35\textwidth]{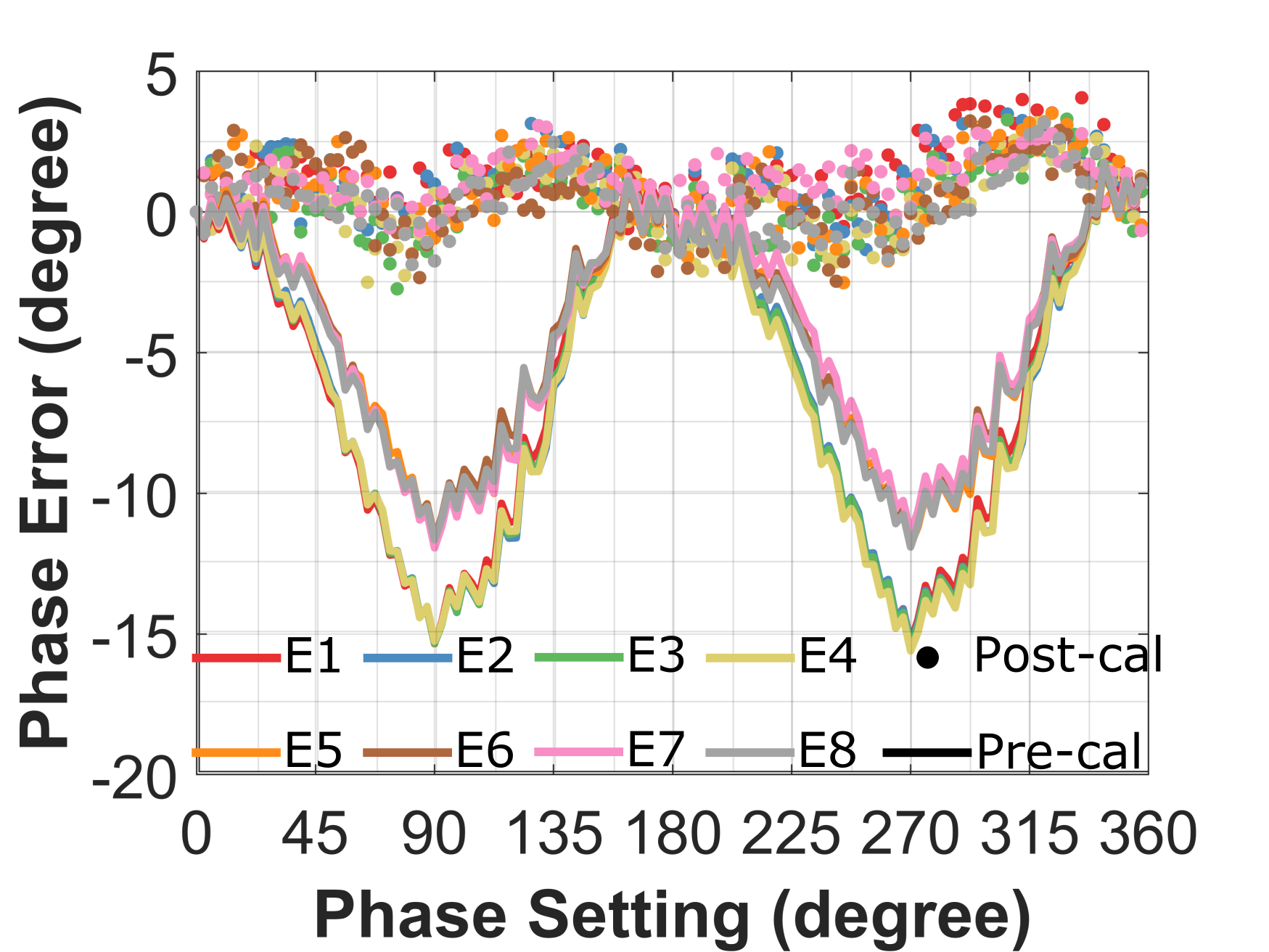}
		\label{fig:POSTPhase18_RX}}

	\caption{VNA measured (a) gain and (b) phase response across phase settings before and after calibration in RX mode.}
	\label{fig:POSTGainRX}
\end{figure}

The same setup is employed in CoMET calibration as in CoMET extraction. Once again, this is run at 6 GHz to demonstrate how CoMET calibration can be used to determine an optimized LUT outside of the intended operating range of the hardware. 
The CoMET calibration algorithm is run to determine the 8x128 LUT for the array. These optimized settings are then applied to the array and it is measured using a VNA. We then compare these VNA measurements to another set of measurements where the array elements are programmed using the original 8 GHz LUT.
The gain and phase data are compared in Fig. \ref{fig:POSTGainTX} and \ref{fig:POSTGainRX} for both TX and RX. 
After calibration, the gain response is equalized across phase settings. The RMS gain error is reduced from 0.8 dB before calibration to 0.1 dB after calibration. Peak-to-peak gain fluctuation is reduced from 4 dB to 0.7 dB. Phase error also improves significantly, however, it is limited by the CoMET accuracy we have already previously demonstrated (2 to 3$^\circ$). After calibration, phase error is 3$^\circ$ in TX mode and 4$^\circ$ in RX mode. At 6 GHz, the RMS phase error is reduced from 8$^\circ$ to 1.4$^\circ$ for TX, and 8$^\circ$ to 1.7$^\circ$ for RX. 

\section{Conclusion} 
Code-modulated embedded test (CoMET) is a technique which can be used for self-test and calibration of phased arrays. CoMET extracts vector information for the array in a parallel fashion using simple test hardware. Previously, CoMET showed larger errors as the phase shift approached the Cartesian axes; however, the rotated axis  methodology is discussed to improve accuracy for this situation.  We experimentally demonstrate CoMET for an eight-element board-level array using ADAR1000 chips from ADI. Measurements indicate that CoMET can extract gain and phase with 0.2 dB and 3$^\circ$ accuracy compared to a VNA. 
Finally, an example calibration method using CoMET is introduced which achieves equalized gain and optimal phase settings for a seven-bit response. In hardware, the root-mean squared gain and phase errors are improved from 0.8 dB and 8$^{\circ}$ before calibration to 0.1 dB and 1.7$^{\circ}$ after calibration.

\section*{Acknowledgment}
This work has been partially supported by Analog Devices Inc. The authors thank Ed Balboni and Bob Broughton from ADI for their guidance and Sandeep Hari from NCSU for help with experiments.

\ifCLASSOPTIONcaptionsoff
  \newpage
\fi

\end{document}